\documentclass[twocolumn,prl,showpacs,preprintnumbers,amsmath,amssymb,showkeys]{revtex4}
\usepackage{graphicx}
\usepackage{dcolumn}
\usepackage{bm}
\usepackage{amsmath}
\usepackage{natbib}
\begin{document}

\preprint{PRL, 2006}

\title{Swinging of red blood cells under shear flow}
\author{Manouk $\rm{Abkarian^{1,3}}$}
\email{abkarian@lcvn.univ-montp2.fr}
\author{Magalie $\rm{Faivre^{2,3}}$}
\author{Annie $\rm{Viallat^{2,3}}$}
\email{viallat@marseille.inserm.fr} \affiliation{\rm{1.} Laboratoire
des Collo\"{i}des, Verres et Nanomat\'eriaux, UMR 5587, CNRS/UM2,
CC26, 34095 Montpellier Cedex 5, France\\ \rm{2.} Laboratoire
Adh\'esion et inflammation, Inserm U600/CNRS UMR 6212 Universit\'e
de la M\'editerran\'ee, Case 937, 163 Av. de Luminy, 13288 Marseille
Cedex 9, France\\ \rm{3.} Laboratoire de Spectrom\'etrie Physique,
UMR 5588 CNRS/UJF, BP 87, 38402 Saint Martin d'H\`eres, France}
\date{\today}

\begin{abstract}
We reveal that under moderate shear stress ($\eta\dot{\gamma}\approx0.1$ Pa) 
red blood cells present an oscillation of their inclination (swinging) superimposed to 
the long-observed steady tanktreading (TT) motion. A model based on a 
fluid ellipsoid surrounded by a visco-elastic membrane initially unstrained 
(shape memory) predicts all observed features of the motion: 
an increase of both swinging amplitude and period (1/2 the TT period) 
upon decreasing $\eta\dot{\gamma}$, a $\eta\dot{\gamma}$-triggered 
transition towards a narrow $\eta\dot{\gamma}$-range intermittent regime of successive 
swinging and tumbling, and a pure tumbling motion at lower $\eta\dot{\gamma}$-values.
\end{abstract}

\pacs{83.50-v; 83.80.Lz; 87.17.Jj} \keywords{tank treading;
tumbling; capsule; membrane elasticity; shape memory}

\maketitle

A human red blood cell (RBC) is a biconcave flattened disk,
essentially made of a Newtonian hemoglobin solution encapsulated by
a fluid and incompressible lipid bilayer, underlined by a thin
elastic cytoskeleton (spectrin network) \cite{Mohandas94}. The
complex structure of RBCs and their response to a viscous shear flow have a
great influence on flow and mass transport in the microcirculation
in both health and disease \cite{Chien87}. The full understanding of
this response requires a direct comprehensive observation of cell
motion and deformation, and a model for deducing the cell
intrinsic properties from its behavior in shear flow. It is
generally admitted that the two possible RBC movements are the
unsteady tumbling solid-like motion \cite{Goldsmith72}, and the
drop-like 'tanktreading' motion for higher shear stresses, where the
cell maintains a steady orientation, while the membrane rotates
about the internal fluid, as reported respectively for RBCs
suspended in plasma or in high-viscosity media and submitted to high
shear stresses \cite{Goldsmith72,Fischer77,Fischer78a,TranSonTay84}.
However, the RBC movement at smaller shear rate and close to the
tumbling-tanktreading transition, has not been fully explored.
Moreover, the actual state of deformation of the elastic skeleton
either in the flowing or in the resting RBC is still conjectural
("shape memory" problem) \cite{Fischer04}. Most models
\cite{TranSonTay84,Sutera89} derive from the analytical framework of
Keller and Skalak (KS) \cite{Keller82}, which treats the RBC as a
fluid ellipsoidal membrane enclosing a viscous liquid. Although this
model qualitatively retrieves the two modes of motion, it does not
capture the observed shear-rate dependency of the
tumbling-tanktreading transition. In particular, the model does not
account for the possible elastic energy storage induced by the local
deformations of the cytoskeleton during tanktreading. Approaches
including membrane elasticity are either restricted to spherical
resting shapes because of analytical complexities
\cite{BarthesBiesel81}, or propose encouraging but still limited
numerical analysis on tanktreading elastic biconcave capsules
\cite{Ramanujan98}.

Here, we reveal a new regime of motion for RBCs under small shear
flow, characterized by an elastic capsule-like oscillation of the
cell inclination superimposed to tank-treading that we name
swinging. We develop a model, which predicts both swinging and the
shear-stress dependency of the tumbling-tanktreading transition. It
demonstrates the existence of the elastic shape memory in the
membrane.

Direct measurements of cell orientation with respect to the flow
direction (angle $\rm{\theta}$) and cell shape (lengths of the long
and small axis of the cell cross-section, $\rm{a_{1}}$ and
$\rm{a_{2}}$ respectively) are provided from side-view microscopic
imaging in a vertical plane parallel to the plane of shear 
\cite{Abkarian05}. We varied the wall shear rate
$\dot{\gamma}$ (in the range 0-5 $\rm{s^{-1}}$) and the outer
viscosity $\eta_{o}$ by suspending RBCs in various solutions of
dextran (concentration 6\%, 7.5\% or 9\% w/w and viscosity 22, 31
and 47 mPa.s respectively). Therefore, the wall shear stress is
varied in a range from 0 to 0.25 Pa.
\begin{figure*}
\includegraphics[width=13.5cm]{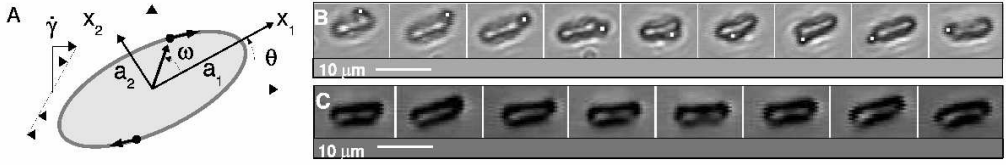}
\caption{\label{fig1} Units
[$\eta_{i}$]=[$\eta_{o}$]=[$\eta_{m}$]=mPa.s, [$\mu$]=Pa and
.[$\dot{\gamma}$]=$\rm{s^{-1}}$. (A) Schematic drawing of a tanktreading ellipsoid in a shear flow. (B)
Rotation of a bead (diameter 1$\mu$m) stuck on the membrane of a RBC
with ($\dot{\gamma}=6$, $\eta_{o}=47$). Time sequence of 1s. (C) RBC
swinging : ($\dot{\gamma}=1.33$, $\eta_{o}=47$). Time sequence of
2s.}
\end{figure*}
For the highest values of the external shear stress
$\eta_{o}\dot{\gamma}$, tanktreading is observed. It is
characterized by i) a quasi-stationary cell shape with insignificant
deformation (maximum variation of $\rm{a_{1}}\leq5\%$), ii) rotation
of the membrane, revealed from the motion of small carboxylated
beads stuck to the membrane (Fig. \ref{fig1}B) and iii) an
oscillation of the cell inclination about a mean value ranging from
$6^{\rm{o}}$ to $25^{\rm{o}}$ (Figs. \ref{fig1}C and \ref{fig2},
see \cite{MovieS1}) at a frequency equal to twice the tanktreading
frequency (Fig. \ref{fig1}B).
\begin{figure}
\includegraphics[width=6.2cm]{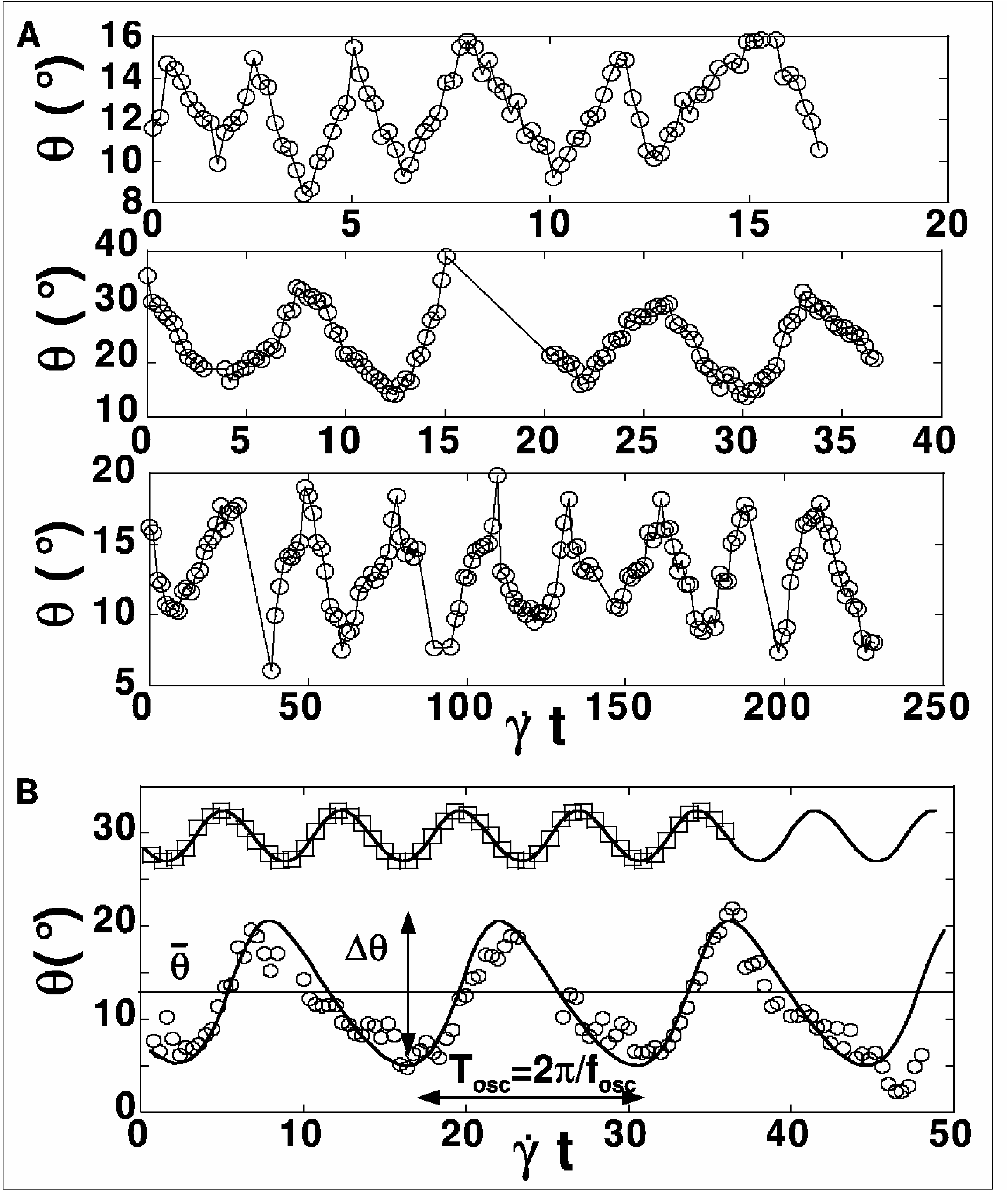}
\caption{\label{fig2} Same units as in Figure 1. (A) Orientation
versus the normalized time $\dot{\gamma}$t for various cells from
top to bottom: ($\dot{\gamma}=1.8$, $\eta_{o}=22$),
($\dot{\gamma}=2.6$, $\eta_{o}=31$), ($\dot{\gamma}=6.6$,
$\eta_{o}=47$). (B) Orientation versus $\dot{\gamma}$t for:
($\circ$) a RBC with ($\dot{\gamma}=0.8$, $\eta_{o}=47$); solid line
from (Eq. \ref{eq2}) with ($\eta_{m}=1120$, $\mu_{m}=0.42$);
($\Box$) a polymeric capsule from \cite{Walter01} with
($\dot{\gamma}=18$, $\eta_{o}=964$). Solid line from (Eqns. \ref{eq2}) with 
the surface moduli ($\eta_{m}.e=0.085$ mPa.s/m, $\mu_{m}.e=0.675$ mPa.m), and
($a_{1} = 278.8 \mu$m, $a_{2} = a_{3} = 170.8$ $\mu$m) obtained from
the size at rest $R_{0} = 224.8$ $\mu$m of the capsule and its mean
deformation during flow: $D=(a_{1}-a_{2})/(a_{1}+a_{2})\approx0.12$
at $\dot{\gamma}=18$.}
\end{figure}
Such characteristics are not seen on tanktreading viscous lipid
vesicles \cite{Abkarian05,Kantsler06}, neither predicted
\cite{Keller82,Kraus96,Noguchi04,Misbah06}. This oscillation is
however observed for non-perfectly spherical elastic
millimeter-scale capsules \cite{Chang93,Walter01} (Fig. \ref{fig2}B)
or protein-coated drops \cite{Emi05}, and in numerical simulations
on biconcave elastic shells \cite{Ramanujan98}. We explain this
phenomenon by assuming RBC shape memory. We state that the local
elements of the composite membrane (cytoskeleton and lipid bilayer),
including the elements which form the rim and the dimples are not
equivalent and are not strained in the biconcave resting shape. They
do consequently not store elastic energy. Thus, during tanktreading,
the elements which form the rim at rest rotate about the stationary
cell shape to reach the dimples after rotation and reciprocally.
These elements are then locally strained and store elastic energy.
Both local deformation and energy storage are periodic: each time
the elements of the membrane make a $\pi$-rotation, they retrieve
their initial shape and are no more strained. We emphasize that 
the periodic storage of energy requires a non spherical unstrained 
state for the RBC. Otherwise, the membrane elements will tanktread 
without modifying the global state of stress of the cell, preserving 
the steady nature of the tanktreading movement. 

\begin{figure}
\includegraphics[width=7cm]{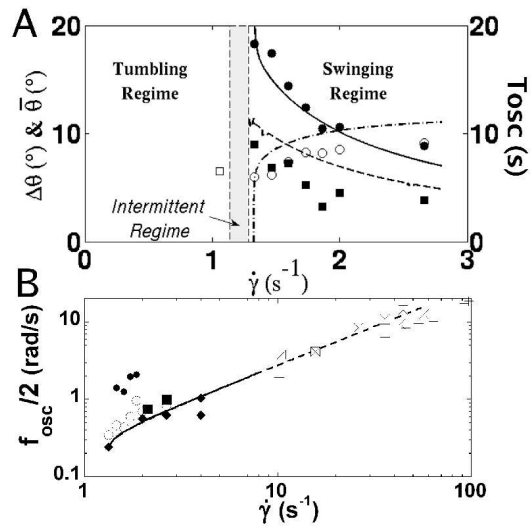}
\caption{\label{fig3} Same units as in Fig. 1. (A) Experimental data on a single RBC
at $\eta_{o}=22$: ($\bullet$) $\Delta\theta$, ($\blacksquare$) $T_{osc}$, ($\square$) 
1 tumbling period value and ($\circ$) $\bar{\theta}$
versus $\dot{\gamma}$. Corresponding curves of the model with 
$\mu_{m}=0.38$ and $\eta_{m}=700$: (\put(0,2){\line(2,0){10}}\hspace{0.4cm}) 
$\Delta\theta$, (- - -) $T_{osc}$ and ($-\cdot-$) $\bar{\theta}$. 
(B) $f_{osc}/2$ versus $\dot{\gamma}$: ($\bullet$) $\eta_{o}=22$; ($\blacklozenge$)
$\eta_{o}=47$; ($\blacksquare$) $\eta_{o}=31$; ($\circ$) Single RBC,
$\eta_{o}=22$. ($\times$) Tanktreading frequencies
$\dot{\omega}$ taken from \cite{TranSonTay84} with $\eta_{o}=35$,
(+) from \cite{Fischer78a} with $\eta_{o}=18$, 31, and 59,
($\boxtimes$) from \cite{Fischer77} with $\eta_{o}=70$. 
The line is a guide for the eyes.}
\end{figure}

In order to derive tractable equations of motion, 
we use the KS model. We consider an oblate ellipsoid filled with a 
viscous liquid and delimited by a
viscoelastic 3D thin membrane, which includes the lipid bilayer and
the underlying cytoskeleton \cite{Ref2D3D}. 
The membrane elements are prescribed to rotate along elliptical trajectories 
parallel to the shear plane, with a linear velocity field \cite{Refvelocity} 
given by: $v_{1}=-\dot{\omega}(a_{1}/a_{2})x_{2}, v_{2}=\dot{\omega}(a_{2}/a_{1})x_{1}, v_{3}=0$, 
where $\omega$ and $\dot{\omega}$ are the phase
angle of a membrane element and its instantaneous frequency of
tanktreading respectively (Fig. \ref{fig1}A).
The KS equation for RBC motion is obtained by
stating that at equilibrium, the total moment exerted by the
external fluid on the cell vanishes (First equation in Eqs.\ref{eq2} below). 
In addition, the movement satisfies the conservation of energy, i.e. the rate of dissipation
of energy in the cell must equal the rate at which work is done by
the external fluid on the cell. KS calculated both rates assuming
viscous energy dissipation in the cell. We add to this latter
contribution the elastic power stored in the periodic elastic strain
of the cytoskeleton \cite{Skotheim06}:
$P_{el}=\int_{\Omega}\rm{Tr({\bf \sigma}:{\bf D})}d\Omega$, where
$\Omega$ is the membrane volume, {\bf D} the eulerian strain
rate tensor derived from the KS velocity field and ${\bf \sigma}$
the shear stress tensor in the membrane; ${\bf\sigma}$ is computed
from the local deformation of the membrane due to tanktreading,
assuming a simple Kelvin-Voigt viscoelastic material: ${\bf
\sigma}=2\eta_{m}{\bf D}+2\mu_{m}{\bf E}$, where ${\bf E}$ is the
Euler-Almansi strain tensor obtained from the KS velocity field.
After some algebra, $\rm{P_{el}}$ writes as
\begin{equation}
\label{eq1} P_{el}=\frac{1}{2}\dot{\omega}
(\frac{a_{2}}{a_{1}}-\frac{a_{1}}{a_{2}})^2
[2\eta_{m}\dot{\omega}+\mu_{m}sin(2\omega)]\Omega
\end{equation}
where $\eta_{m}$ and $\mu_{m}$ are the membrane viscosity and the
shear modulus respectively. Conservation of energy provides a
constraint on the allowable RBC motion and yields a second
differential equation (for more details see \cite{MovieS1}). 
The two coupled equations are:
\begin{eqnarray}
\begin{split}
\label{eq2}
\dot{\theta} & =-(\frac{1}{2}\dot{\gamma}+\frac{2a_{1}a_{2}}{a_{1}^2+a_{2}^2}\dot{\omega})+\frac{1}{2}\dot{\gamma}\frac{a_{1}^2-a_{2}^2}{a_{1}^2+a_{2}^2}cos(2\theta)\\
\dot{\omega} & =-\frac{\eta_{o}f_{3}\dot{\gamma}}{\eta_{o}f_{2}-\eta_{i}(1+\frac{\eta_{m}}{\eta_{i}}\frac{\Omega}{V})f_{1}}cos(2\theta)\\
             & + \frac{\frac{1}{2}f_{1}\mu_{m}\frac{\Omega}{V}}{\eta_{o}f_{2}-\eta_{i}(1+\frac{\eta_{m}}{\eta_{i}}\frac{\Omega}{V})f_{1}}sin(2\omega)
\end{split}
\end{eqnarray}
where $\dot{\theta}$ is the time derivative of the cell inclination,
$f_{1}$, $f_{2}$ and $f_{3}$ are geometrical constants and V is the
RBC volume (same definition as in \cite{Keller82}). The limiting
case $\mu_{m}=\eta_{m}=0$ corresponds to KS. We numerically solved
the equations using the following set of parameters for RBCs :
$\rm{a_{1} = a_{3}} = 4 \mu$m, $\rm{a_{2}} = 1.5 \mu$m, $\Omega =
\Sigma.e$ \cite{Note02}, where $\Sigma$ is the oblate ellipsoid area
and e=50 nm is the membrane thickness \cite{Heinrich01}. $\eta_{i}$
is fixed at the physiological value of $10$ mPa.s
\cite{TranSonTay84} and $\eta_{m}$ is adjusted in the range 0.7-2
Pa.s \cite{TranSonTay84}. We obtain $\theta(t)$, $\dot{\theta}(t)$,
$\omega(t)$ and $\dot{\omega}(t)$. Suitable couples of $\mu_{m}$ and
$\eta_{m}$ values were found to reproduce experimental RBC oscillations as
seen in Fig. \ref{fig2}B (see \cite{MovieS1}). One example of experimental 
capsule oscillation extracted from \cite{Walter01} is
presented as well in Fig. \ref{fig2}B. An insight of experimental and numerical
swinging curves is provided from three parameters: the magnitude
$\Delta\theta=\theta_{max}-\theta_{min}$, the mean angle $\bar{\theta}$ and the period $T_{osc}$ (or the frequency $f_{osc}$)
of oscillation (Fig. \ref{fig2}B). The $\dot{\gamma}$-variations of these parameters 
are illustrated for one red blood cell in Fig. \ref{fig3}A together with a numerical 
solution of Eqns \ref{eq2}. While $\bar{\theta}$ decreases for decreasing $\dot{\gamma}$ 
down to 0, $T_{osc}$ and $\Delta\theta$ increase until $\dot{\gamma}$ reaches a 
critical value ${\dot{\gamma}_{c}^{-}}$ below which the cell tumbles at least
once. Besides direct observations (Fig. \ref{fig1}B), we illustrate the factor 2 of proportionality between $f_{osc}$ 
and $\dot{\omega}$ which relates the movement of swinging of a cell to the movement of tanktreading 
of its membrane, by reporting on a same graph in Fig. \ref{fig3}B, variations of $f_{osc}/2$ and $\dot{\omega}$ 
versus $\dot{\gamma}$, for different swinging RBCs observed in our experiment and for different tanktreading RBCs 
observed in the litterature at higher $\dot{\gamma}$ \cite{Fischer78a,TranSonTay84}. Fig. \ref{fig3}B shows 
indeed the continuity of the two regimes even though measured on different type of movement. Therefore, all experimental 
characteristics are well captured by the model. Indeed, by treating the elastic contribution as a small 
perturbation in the second equation of (2), valid in the linear part, one recovers the steady KS solution at the
order 0 of the parameter in front of $sin(2\omega)$, while at the
first order one finds that $\theta$ oscillates at twice the
tanktreading frequency (linear in $\dot{\gamma}$) and $\Delta\theta$ scales as
($\mu_{m}/\dot{\gamma}$).

\begin{figure}
\includegraphics[width=7.5cm]{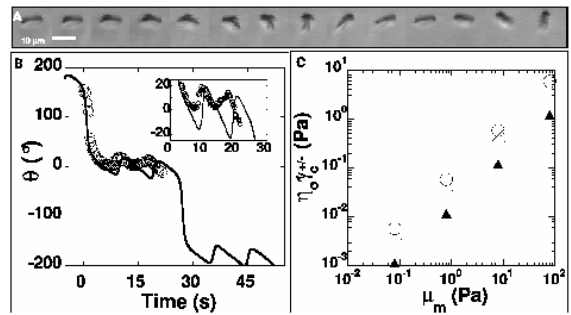}
\caption{\label{fig4} Same units as in Fig. 1. (A) The transition
from swinging to tumbling induced by decreasing $\dot{\gamma}$ is
associated with a transient localized deformation ($\eta_{o}$=47,
$\dot{\gamma}$=2.66). Time sequence of 1s. (B) ($\circ$) Successive
swinging and tumbling at ($\eta_{o}=22$, $\dot{\gamma}$=1.526);
($---$) numerical calculus with ($\eta_{o}=22$,
$\dot{\gamma}$=1.526, $\mu_{m}=0.454$, $\eta_{m}=700$). (C)
Theoretical shear-stresses of transition versus $\mu_{m}$ with
$\eta_{m}=1000$: ($\circ$) $\eta_{o}{\dot{\gamma}_{c}^{-}}$;
($\times$) $\eta_{o}{\dot{\gamma}_{c}^{+}}$ and ($\blacktriangle$)
$\eta_{o}(\dot{\gamma_{c}}^{-}-\dot{\gamma_{c}}^{+})$.}
\end{figure}

Contrarily to KS prediction, we observe that the transition of
movement from tumbling to tanktreading (respectively tanktreading to
tumbling) is induced by tuning up (respectively down) the applied
shear rate. This transition is illustrated in Fig. \ref{fig4}A.
Its more striking feature, predicted and experimentally observed is
the existence of a regime of movement where the cells present
successively swings and tumbles at a given $\dot{\gamma}$ (Fig.
\ref{fig4}B). The model gives the $\dot{\gamma}$-range
($[\dot{\gamma}_{c}^{+},\dot{\gamma}_{c}^{-}]$) where this regime
exists. Given the experimental constraints, it is not easy to follow
the cells sufficiently long to observe a large series of tumbles and
swings. We define over a time scale of $\sim20$ s, the shear rate
corresponding to a change in movement from swinging to tumbling
($\dot{\gamma}_{c}^{<}$) with decreasing $\dot{\gamma}$, and from
tumbling to swinging ($\dot{\gamma}_{c}^{>}$) with increasing
$\dot{\gamma}$. The difference $\dot{\gamma}_{c}^{>}$ -
$\dot{\gamma}_{c}^{<}$, that we call hysteresis (for instance for
two different RBCs at $\eta_{o}$ =31 mPa.s : ${\dot{\gamma}_{c}^{<}}
= 0.47$ s$^{-1}$  , ${\dot{\gamma}_{c}^{>}} = 1$ s$^{-1}$ and
${\dot{\gamma}_{c}^{<}} = 0.8$ s$^{-1}$, ${\dot{\gamma}_{c}^{>}} =
1.73$ s$^{-1}$ respectively) has the same order of magnitude than
the theoretical $\dot{\gamma}$-range
$[\dot{\gamma}_{c}^{+},\dot{\gamma}_{c}^{-}]$ of intermittency.
Finally as it is seen by requiring the second term in the second
equation of (2) to be of the same order of magnitude as the first part, the
critical shear rate should scale as $\mu_{m}/\eta_{o}$. It is indeed
numerically observed (Fig. \ref{fig4}C). Both critical values of
$\eta_{o}\dot{\gamma}_{c}^{+}$ and $\eta_{o}\dot{\gamma}_{c}^{-}$
are mainly governed by the RBC elastic contribution for given cell geometry and 
may provide an average determination of $\mu_{m}$ by observing a large sample of RBCs (Fig. \ref{fig4}C).
\begin{figure}
\includegraphics[width=5cm]{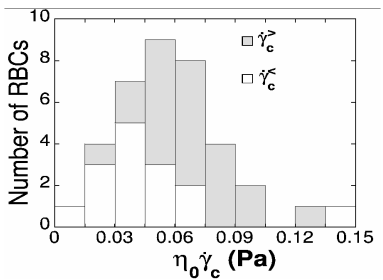}
\caption{\label{fig5} Distribution function of critical shear
stresses of transition for increasing ($\dot{\gamma}^{>}$) and
decreasing ($\dot{\gamma}^{<}$) shear rates.}
\end{figure}
Distribution functions of $\eta_{o}\dot{\gamma}_{c}^{>}$ and
$\eta_{o}\dot{\gamma}_{c}^{<}$ measured on a large RBC sampling are
shown in Fig. \ref{fig5}. They characterize the natural variability
of the RBC elastic modulus and illustrates the additional hysteresis
effect. From Fig. \ref{fig5} and Fig. \ref{fig4}C, we find that
$\mu_{m}$ ranges in the interval 0.14-2 Pa. By setting the 2D shear
modulus $\mu_{m,S}=\mu_{m}.e$ we obtain values ranging from 0.07 to
1$\times10^{-7}$ N/m ($e = 50$ nm) below that usually reported
\cite{Mohandas94,BarthesBiesel81}. We also find a comparable
difference on $\mu_{m,S}$ with that reported for the elastic
capsules by Walter et al \cite{Walter01}. This underestimation
likely originates from the major simplifications we made in order to
obtain simple analytical equations allowing the full understanding
of the physics of the problem: i) simplistic constitutive equations,
ii) approximate KS velocity field, which may overestimate membrane
deformations. In particular, Tran son Tay et al \cite{TranSonTay84}
suggested that the Secomb-Skalak area conserving velocity field
\cite{Secomb82} would lead to a 70\%-increase of the membrane
viscosity compared to that derived from the KS model. iii) treatment
of deformations from a 3D description of the membrane of RBCs and
capsules although these systems form mainly 2D-shells \cite{Ref2D3D}. 
However, the main interest of this tractable model is to understand 
the role of the various physical parameters on the movement.
For example, for given external viscosity and shear rate, $\Delta\theta$ is not
much sensitive to values of $\eta_{i}$ and $\eta_{m}$ taken in the
physiological range. $\Delta\theta$ is essentially fixed by $\mu_{m}$, 
and thus the measurement of the amplitude of swinging as a function of the shear stress 
may provide a complementary method to accurately determine $\mu_{m}$ on single flowing 
RBCs in a given sample. 

In conclusion, the swinging movement and the shear-stress triggered
transition of motion of RBCs demonstrate the existence of their
shape memory and is a signature of their membrane shear elasticity.
Despite its simplicity, our model provides a good description of the
observed behaviors and we believe that a more sophisticated model
should allow an easy and sensitive determination of individual RBC
mechanical properties. Finally, such experimental approach coupled 
to the refined model might be applied to a wide variety of 
soft shells \cite{Chang93,Walter01,Emi05} and holds promises for
applications in surface rheology measurements.

We would like to thank Dr. J. Skotheim for discussions, B.
Carpentier for experimental help, Pr. H. A. Stone for fruitful
discussions and Pr. H. Rehage for providing us with the data on the
capsule used in Fig. 2B.


\begin{thebibliography}{26}
\bibitem{Ref2D3D}
Because of the 3D nature of the KS velocity field, 
the membrane deformation is considered to be 3D while other non algebraically 
tractable approaches considered the deformation field to be only 2D \cite{TranSonTay84,Secomb82}.
\bibitem{Refvelocity}
The KS velocity field is not area conserving. 
However improved models \cite{Secomb82} that account for incompressibility 
are much more complex algebraically and it is generally
considered that the simple KS velocity field provides a good approximation
of the membrane deformation \cite{TranSonTay84}.
\end{thebibliography}
\end{document}